\RequirePackage{fix-cm}
\documentclass[twocolumn]{svjour3}          
\usepackage{graphicx,color}

\begin{document}


\title{S. Vonsovsky and the dawn of the theory of strongly correlated systems}


\titlerunning{S. Vonsovsky and the dawn of the theory of strongly correlated systems}        

\author{V. Yu. Irkhin        
	 \and
        S.V. Streltsov 
}


\institute{V.Yu. Irkhin and S.V. Streltsov\at
              M. N. Mikheev Institute of Metal Physics, 620108 Ekaterinburg, Russia \\
              \email{valentin.irkhin@imp.uran.ru}           
}

\date{Received:  \today / Accepted: date}

\maketitle

\begin{abstract}
A survey  of scientific and social activity of S. V. Vonsovsky is presented. His fundamental works on the theory of electronic properties and magnetism of transition and rare-earth metals and their compounds are reviewed. The historical development of the many-electron polar  (preceding the famous Hubbard model) and $s-d(f)$ exchange (``Vonsovsky-Zener'', ``Kondo lattice'') models is considered.

\keywords{magnetism \and electron correlations \and polar model \and $s-d(f)$ exchange model \and Kondo effect}
 \PACS{71.27.+a \and 75.10.Lp \and 71.30.+h}
\end{abstract}

Sergey Vasiljevich Vonsovsky (1910-1998) was one of founders of the Institute of Metal Physics in Sverdlovsk (now Yekaterinburg), as well as the entire Ural scientific school on the quantum theory of solids and the physics of magnetic phenomena. Most of the Ural theoretical physicists of the last century considered him as their teacher. 

Vonsovsky  was the founder of the famous ``Kourovka'' - the winter school of theoretical physics, which has been held in the Urals since 1960 and is known for its warm and democratic atmosphere. This volume of the Journal of Superconductivity and Novel Magnetism represents contributions of past and present Kourovka attendees.  


In the present paper, we discuss most important scientific works by Vonsovsky, especially in the many-electron theory of solids (together with his teacher Semen Petrovich Shubin). These contributions are not very well known to a younger generation grown on articles by J. Hubbard and P. W. Anderson (older colleagues are of course familiar with them), while ideas introduced by S. Vonsovsky into the field of strongly correlated electron systems were nonetheless seminal. This is one of the aims of the present paper - to shortly remind about two of his models: the polar (Section \ref{Polar}), and $s-d(f)$ exchange (Section \ref{s-d}) models. Some historical background is presented in more detail in Section~\ref{life}.

\section{Polar model \label{Polar}}

In their works on the polar model \cite{1,2}, Shubin and Vonsovsky stated a goal to simultaneously describe a wide range of phenomena in solids, including magnetism and electrical conductivity. Already in 1934-1935 (i.e. nearly 30 years before Hubbard) they formulated a minimal model for interacting electrons in a solid. This so-called polar model was proposed as a synthesis of the Heisenberg homeopolar model describing the system of localized moments, and the Slater method for describing the itinerant-electron system of a metal. In a crystal with one electron per atom, it means taking into account polar states –- doubles and holes, i.e. doubly occupied and empty sites, and leads to a physically transparent picture. In the original formulation of the model, hoppings of electrons from site to site were taken into account, as well as all types of interelectron  interaction.
These papers, which provided the foundations of many-electron solid-state physics, were carried out by means of difficult and cumbersome calculations (there was no convenient method of second quantization at that time).

The polar model was further generalized by Bogolyubov \cite{bog}, who derived its Hamiltonian via the successive expansion in terms of the hopping parameters (loosely speaking the overlap between atomic wave functions) in the second-quantization representation. In the simplest case of a nondegenerate band, this Hamiltonian can be written in the form
\begin{eqnarray}
	{\mathcal H} &=&\sum_{\nu _1\neq \nu _2,\sigma }t_{\nu _1\nu _2}c_{\nu _1\sigma
	}^{\dagger }c_{\nu _2\sigma }^{} \nonumber\\
&+&\frac 12\sum_{\nu _i\sigma _1\sigma _2}I_{\nu _1\nu _2\nu _3\nu
		_4}c_{\nu _1\sigma _1}^{\dagger }c_{\nu _2\sigma _2}^{\dagger }c_{\nu
		_4\sigma _2}^{ }c_{\nu _3\sigma _1}^{ }.
	\label{eq:G0}
\end{eqnarray}
Here, $c_{\nu\sigma}^{\dagger }$ are the creation operators for an electron at the site $\nu$ with the spin projection $\sigma$, $t_{\nu _1\nu _2}$ are the hopping integrals (matrix elements of
one-electron transfer). Thus, first term is the kinetic energy; being transformed to the reciprocal space it leads to formation of the band structure. The second term in (\ref{eq:G0}) describes interaction between electrons set up by the matrix elements $I_{\nu _1\nu _2\nu _3\nu_4}$. One can introduce very different interactions in this way, e.g. $V_{\nu _1\nu _2}=I_{\nu_1\nu_2\nu_1\nu_2}$ is the Coulomb interaction at different sites (responsible, for example, for the charge ordering), $J_{\nu _1\nu _2}=-I_{\nu_1\nu_2\nu_2\nu_1}$ is the ``direct'' exchange interaction (this term is obtained by the transposition (exchange) of spin subscripts).

\begin{figure}
	\includegraphics[width=0.49\textwidth]{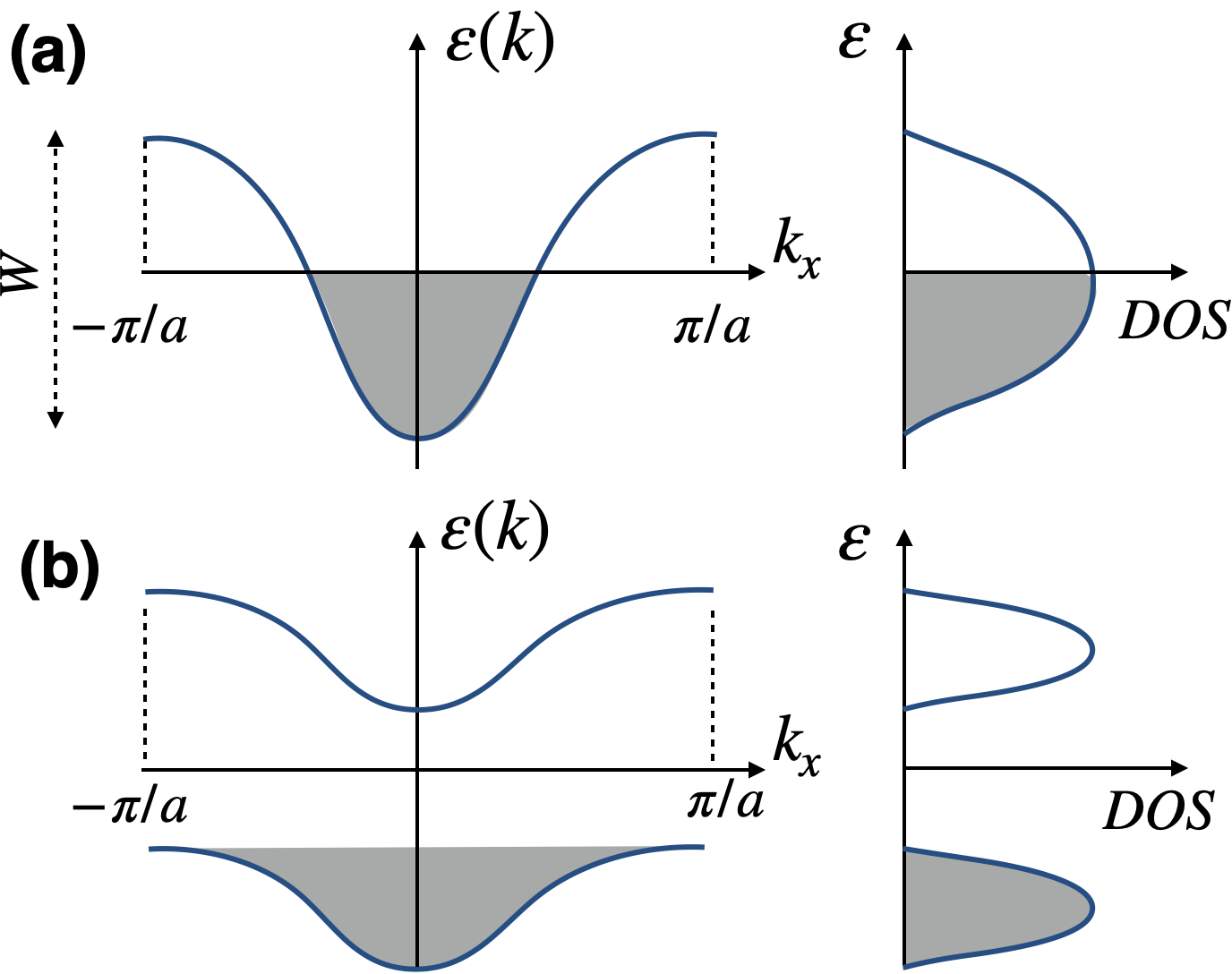}
	\caption{The band structure due to the first term in (\ref{eq:G.1}) is shown in (a). (b) illustrates formation of lower and upper Hubbard bands in regime of large on-site repulsion between electrons. The one-dimensional lattice with period $a$ is used for this sketch.}
	\label{Hubbard}       
\end{figure}

The next step was made by Gutzwiller \cite{Gutzwiller} and Hubbard \cite{H1,H3,H4}, who singled out in the model the most important part of the Coulomb interaction -- the strong repulsion of electrons at the same site $U=I_{\nu\nu\nu\nu}$. In the case of the non-degenerate band,  the Hubbard Hamiltonian is given by
\begin{equation}
	\mathcal{H}=\sum_{\mathbf{k}\sigma
	}\varepsilon_{\mathbf{k}}c_{\mathbf{k}\sigma }^{\dagger }c_{\mathbf{k}\sigma
	}+ U\sum_ic_{i\uparrow }^{\dagger
	}c_{i\uparrow }c_{i\downarrow }^{\dagger }c_{i\downarrow }, \label{eq:G.1}
\end{equation}
where $ \varepsilon_{\mathbf{k}}$  is the band spectrum. The Hubbard model was widely used for studying ferromagnetism of itinerant electrons, metal-insulator transition, and other physical phenomena connected with electron correlations.  In spite of the obvious simplicity, this model retains very rich physics. Its strict treatment is an extremely complicated and still not solved (in a general case) problem. Here we only mention that the first term results in formation of the bands with the width  $W$. If the repulsion between electrons described by the second term is large, $U > W$, there appear the lower and upper Hubbard bands and the system becomes insulating, as illustrated in Fig.~\ref{Hubbard}, for a detailed consideration we refer to a more specialized literature, e.g., \cite{Fazekas,Montrosi,II1,Kivelson}.

In Ref.\cite{H4} Hubbard introduced the many-electron representation of  generalized projection operators:
\begin{equation}
	X(\Gamma ,\Gamma ^{\prime })=|\Gamma \rangle \langle \Gamma
	^{\prime }|, \quad \sum_\Gamma X(\Gamma ,\Gamma )=1,
	\label{eq:A.22}
\end{equation}
where  $|\Gamma \rangle $  are the exact (many-body) eigenstates of the local part of the Hamiltonian (on-site interaction and the crystal-field splitting  in the case of a few orbitals), see also Refs. \cite{II,Ovchinnikov}. The Hubbard operators describe transitions between different many-electron terms.

Let us consider what kind of {\it local} states should be included into the basis set to solve the non-degenerate model (\ref{eq:G.1}) at half-filling, i.e. with one electron per site in average. Even in this situation there are not only states with one electron having spin up $| + \rangle$ or down $| - \rangle$ -- so-called non-polar states (no additional electrons or holes), but also polar states without electrons $| 0 \rangle$ (hole) and with two electrons $| 2 \rangle$ (double) at a given lattice site.  Shubin and Vonsovsky argued that the correct description should include not only one-electron Bloch states obtained by Fourier transformation of  $| + \rangle$ and $| - \rangle$, but also polar states. This is exactly how correlation effects appear.  

Shubin and Vonsovsky used in their works a quasi-classical approximation. It in fact reduces to the replacement of $X$-operators by c-number functions $\varphi_{i}, \psi _{i}, \Psi _{i}$, and $\Phi _{i}$, which determine the probability amplitude of the $i$th site occupation by zero, one or two electrons
\begin{eqnarray}
	X_{i}(+,0)\rightarrow \varphi _{i}^{\ast }\Psi _{i}, X_{i}(2,-)\rightarrow
	\Phi _{i}^{\ast }\psi _{i}, \nonumber 
	X_{i}(2,0)\rightarrow \Phi _{i}^{\ast }\Psi _{i}
	\label{quasicl}
\end{eqnarray}%
with a subsidiary normalization
\begin{equation}
	|\varphi _{i}|^{2}+|\psi _{i}|^{2}+|\Phi _{i}|^{2}+|\Psi _{i}|^{2}=1.
\end{equation}%
In this approximation the total energy can be calculated via the variational principle with a wave function
\begin{equation}
	\phi =\prod_{i}\left(\varphi _{i}^{\ast }X_{i}(+,0)+\psi _{i}^{\ast
	}X_{i}(-,0)+\Phi _{i}^{\ast }X_{i}(2,0)+\Psi _{i}^{\ast } \right)|0\rangle.
	\label{dops}
\end{equation}
This function mixes the Fermi- and Bose-type excitations and thereby does not satisfy the Pauli principle. Nevertheless, the quasi-classical approximation provides a rough description of some physical phenomena, e.g., of metal-insulator transition in spirit of the Gutzwiller approximation \cite{Gutzwiller} (see Ref. \cite{7}).

Further on, the many-electron theory was developed by introducing various representations of auxiliary (``slave'') Bose and Fermi operators for the $X$-operators (see, e.g., \cite{i}). In particular, in connection with the theory of high-temperature superconductivity, P.W. Anderson \cite{633b} put forward the idea to separate spin and charge degrees of freedom of an electron by using the representation
\begin{equation}
	c_{i\sigma }^{\dagger
	}=X_i(\sigma ,0)+\sigma X_i(2,-\sigma )=s_{i\sigma }^{\dagger
	}e_i+\sigma d_i^{\dagger }s_{i-\sigma }.
	\label{eq:6.131}
\end{equation}
Here $s_{i\sigma }^{\dagger }$ are the creation operators for neutral fermions (spinons), and $e_i^{\dagger }$, $d_i^{\dagger }$ for charged spinless bosons. The physical meaning of
such exotic excitations with unusual statistics can be explained as follows. Let us consider a lattice with one electron per site with a strong Hubbard repulsion $U$, so that each site is neutral.
In the ground resonance valence bond (RVB) state, each site participates in one bond. When the bond is broken, there appear two uncoupled sites, each of them possessing spins 1/2. The corresponding excitations (spinons) are not charged. At the same time, the vacancy (hole) in the system carries a charge, but not a spin.

\section{$s-d(f)$ exchange model \label{s-d}}

In contrast to the polar model, in the $s-d(f)$ exchange model \cite{sd,sd1} the existence of localized magnetic moments is not derived, but postulated; in this case, the subsystems responsible for magnetism and electrical conductivity are separated: itinerant ``$s$-electrons'' determine transport properties, and localized ``$d$-electrons'' are the source of the magnetic moments. Initially,  the $s-d$ exchange model was constructed for $d$-metals, where it gives only a semi-phenomenological picture, but its postulates turned out to be even better justified for systems with $f$-electrons, which are usually well localized. Later, the $s-d$ model was used to describe a number of $d$- and $f$-compounds, for example,  so-called ``heavy fermion systems'' with giant electronic specific heat. It also turned out to be surprisingly fruitful for magnetic semiconductors. 

\begin{figure}
	\includegraphics[width=0.49\textwidth]{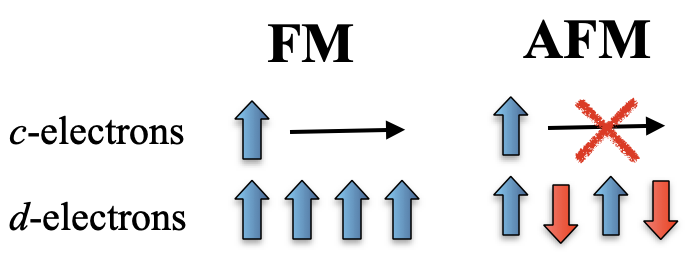}
	\caption{Sketch illustrating the Zener's double-exchange mechanism of ferromagnetism (FM) based on the idea of itinerant ($s$) and localized ($d$) electrons.}
	\label{DE}       
\end{figure}

In the western papers, the $s-d$ model is sometimes called the Vonsovsky-Zener model in the honor of the American scientist C. Zener, who in 1951 proposed \cite{Zener} the double-exchange mechanism of ferromagnetism. Basically it was based on the same conception of two very different  types of electrons. Localized $d$-electrons provide their spins, which can be ferro- or antiferromagnetically ordered, but contact (in fact, Hund's intra-atomic) exchange prefers ferromagnetic order, since in this situation $c$ electrons propagate easily  through the lattice as shown in Fig.~\ref{DE}, see also, e.g., Ref.\cite{Khomskii}. Such concepts were used to theoretically describe giant magnetoresistance in manganites based on LaMnO$_3$.

The Hamiltonian of the $s-d (f)$ model  can be presented as
\begin{eqnarray}
	\mathcal{H}= \sum_{\mathbf{k}\sigma \nonumber
	}t_{\mathbf{k}}c_{\mathbf{k}\sigma }^{\dagger }c_{\mathbf{k}\sigma
	}+\sum_{\mathbf{q}}J_{\mathbf{q}}\mathbf{S}_{-\mathbf{q}}
	\mathbf{S}_{\mathbf{q}}\\
	-I\sum_{i\sigma \sigma ^{\prime
	}}(\mathbf{S}_i\mbox{\boldmath$\sigma $}_{\sigma \sigma ^{\prime
	}})c_{i\sigma }^{\dagger }c_{i\sigma ^{\prime }}, \label{eq:G.2}
\end{eqnarray}
where $\boldmath \sigma $ stands for the Pauli matrices, $\mathbf{S}_i$ are operators of localized spins, summation in the last term runs over sites and spins, $I$ is the parameter of the $s-d(f)$ exchange interaction, which is assumed to be contact. Often
(for example, in rare-earth metals), the interaction between the localized spins $J_{\bf{q}}$ is in fact the indirect Ruderman-Kittel-Kasuya-Yosida (RKKY) interaction, see e.g., Ref.~\cite{Yosida}, via conduction electrons, which appears in the second order of the perturbation theory in $I$. However, when constructing perturbation theory, it is convenient to include $J$ in the zero Hamiltonian.

The fate of the $s-d$ model turned out to be very interesting and difficult, in a sense even dramatic. In the modern physical literature the $s-d$ model is frequently called the Kondo (or Kondo lattice) model. This  historical injustice is connected with an exceptional importance of the Kondo effect \cite{kondo}, which is, maybe, the most beautiful phenomenon in condensed matter physics. 
However, Kondo in its pioneering study of the anomalous magnetic $s-d (f)$ scattering proceeded just  from the second-quantization representation Hamiltonian, which was proposed by Vonsovsky and Turov \cite{sd1}. In fact, the field of application of the $s-d$ model goes far beyond the systems with a considerable ``Kondo'' suppression of localized moments: this describes successfully many magnetic materials.

Unlike the intra-atomic Coulomb (Hubbard) interaction, the $s-d(f)$ interaction $I$ is, as a rule, not
strong. Nevertheless, it leads to essential effects in the electronic spectrum. From a microscopic viewpoint, the $s-d(f)$ interaction can have different nature. In rare-earth metals and in a number of their  insulating and semiconducting compounds this is the intra-atomic Hund type exchange, which is ferromagnetic. In metallic systems, frequently the $s-d(f)$ exchange interaction is an effective one, caused by the hybridization between the conduction $s$-bands and atomic levels of the $d (f)$-electrons. In this case, the sign of $I$ is negative (antiferromagnetic).  

Such an important physical phenomenon as the Kondo effect appears in higher orders of perturbation theory: there occurs a singular logarithmic correction effective (renormalized) coupling parameter in the second order, and the corresponding contribution to resistivity in the third order. Thus, in the case of a negative (antiferromagnetic) exchange $I$, the effective exchange interaction becomes infinite (the infrared divergence leading to anomalies of electron properties), so that the magnetic scattering leads to a complete screening of localized moments \cite{558}.

Being formulated already in the first half of XX century, the polar and $s-d(f)$ exchange models are still used in different forms in the modern condensed matter physics. They provide a basis for new theoretical concepts describing physical phenomena (including exotic ones) discovered by experimentalists. The model approaches which treat effects of strong electron correlations in $d$- and $f$-compounds turn out to be very useful from the point of view of the qualitative microscopic description and provide beautiful physics.

\section{Historical background \label{life}}
S.V. Vonsovsky was born and spent his childhood in Tashkent. His mother Sofya Ivanovna came from an old, but impoverished noble family; maternal grandfather was a country doctor. Father, Vasily Semenovich, originated from a large peasant family. After graduating from the gymnasium and the Faculty of Physics and Mathematics of Moscow University, he went to teach children physics and mathematics in Turkestan. Here he became the director of the Tashkent Women's Gymnasium.  Before the revolution, he was an active member of the Constitutional Democratic Party of People's Freedom (Cadets), being personally acquainted with the leader of this party, Petr Milyukov.

Sofya Ivanovna, being an excellent pianist, thought of making her son a musician. Having retained his love for classical music for life, young Sergey nevertheless followed the path of his father, choosing physics, and never regretted it. Being already a famous scientist, he liked to repeat ``Physics is the mother of all sciences''.

S.V. Vonsovsky graduated from Leningrad University in 1932. He gratefully recalled V.I. Smirnova, O.D. Khvolson, Yu.A. Krutikova, V.A. Fock, P.I. Lukirsky and his other teachers at the Leningrad State University. After graduation, he was sent to the Ural Physical-Technical Institute (later Institute of Metal Physics) in Sverdlovsk, where he began work under the guidance of a young professor Semen Petrovich Shubin (1908-1938), who became his main teacher. 

\begin{figure}
	\includegraphics[width=0.49\textwidth]{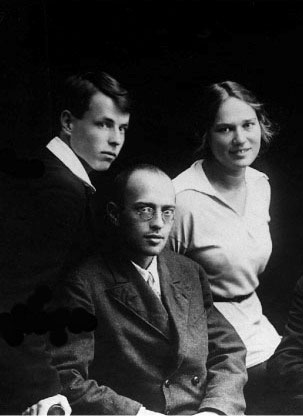}
	\caption{Sergey Vasiljevich Vonsovsky, Semen Petrovich Shubin, Lubov Abramovna Shubina}
	\label{fig:1}       
\end{figure}

Soon there appeared the articles by Shubin and Vonsovsky on the polar model of metals and semiconductors \cite{1,2}.  They were printed in the prestigious English journal Proceedings of the Royal Society and (in more detail) in the Kharkov journal Phys. Zs. UdSSR published in German. Unfortunately, these paper remained largely underestimated, being in a way ahead of their time. The ideas contained in them were then rediscovered and developed abroad (here we should mention again the famous Hubbard model, which is a special case of the polar model).

In 1937, Shubin was arrested for his previous participation in the Trotskyist opposition to Stalinism and  died in the Kolyma Gulag camps. Vonsovsky, who lived in the same house as Shubin, was a witness at his last arrest. Shubin had two children, and the third child was to be born soon. Vonsovsky took upon himself all the worries about the family of an untimely deceased friend and teacher. Lubov Abramovna Shubina (Shatskina) was sister of Lazar Abramovich Shatskin (1902-1937) -- one of the founders of Komsomol, which was also arrested and killed for his struggle against Stalinism.

It would hardly have been possible to physically survive without the support of M.N. Mikheev, a long-term director of the Institute of Metal Physics, who himself did not escape temporary dismissal for a his principled stand. Vonsovsky was fired from the Institute too, but his colleague Rudolf Janus helped by employing him to an experimental laboratory for several months. 

Since 1939 S.V. Vonsovsky is head of the Department of Theoretical Physics, later deputy director of the institute. In wartime, Vonsovsky together with Rudolf Janus carried out defense orders related to the flaw detection of shells, and worked in Nizhny Tagil. After the war, scientific activity gradually improved. In addition, Vonsovsky had to pay much attention to administration. With his active participation there began intensive studies on neutronography, low-temperature physics, radiation physics, and the theory of dislocations in the institute. 

Vonsovsky's works in the field of the quantum theory of solids, especially the many-electron theory of metals and semiconductors, the theory of ferro- and antiferromagnetism, and superconductivity, received worldwide recognition. In 1946, he proposed the $s-d$ exchange model \cite{sd,sd1} partly based on the work with Shubin. Vonsovsky wrote:

``In parallel with the polar model, there was development  of another -- now so-called $s-d$- or $s-f$-exchange model of transition metals, which was invented by Semen Petrovich during his oral discussion with L.D. Landau (S.P. greatly respected and appreciated him). I finished developing this model with my employees when Semen Petrovich was no longer with us''.

Together with his students and colleagues Vonsovsky elaborated the theory of ferromagnetism of alloys, magnetic anisotropy and magnetostriction. His papers were also related to the theory of superconductivity in transition metals and alloys, the problem of the coexistence of ferromagnetism and superconductivity. The fundamental monograph ``Magnetism'' \cite{magn} has become a reference book for physicists of several generations and has been translated into many languages. Other monographs by Vonsovsky \cite{shur,res,el,iz,vk} are widely known in Russia and abroad.

Since 1947, Vonsovsky became a professor at the Ural State University, where he lectured for many years quantum mechanics, solid state physics, and the theory of magnetism; for a number of years he headed the Department of Theoretical Physics. Being a leading magnetologist in the country,  Vonsovsky was chairman of the Scientific Council on Magnetism at the USSR Academy of Sciences for thirty years. Many major Soviet and international conferences on magnetism (including the well-known Moscow State University conferences) were organized under his leadership. He closely communicated with many prominent foreign scientists. From the moment of the formation of the Ural Scientific Center of the USSR Academy of Sciences (1971) to 1985, Vonsovsky served its chairman.

Physics of Metals and Metallography journal was established by Vonsovsky in 1955 (he was its permanent editor-in-chief) and became popular in the physical society of USSR. The scientific and social activities of Vonsovsky were estimated by many government awards, including the Orders of Lenin (three times),  of the Red Star, of the Red Banner of Labor (twice) and the title of Hero of Socialist Labor (1969). He was a laureate of the State Prize (1975, 1982), received Vavilov medal (1982), Demidov Prize (1993), was a member of the German and Polish Academies of Sciences.  Vonsovsky was a deputy of the Supreme Soviet of the Russia Federation (1963-1971).

The most important personal quality Vonsovsky was kindness: he was simply physically incapable of hurting a person by refusing when he was asked for something. In Soviet times, he helped his colleagues and students a lot, not only contributing to the solution of material problems, but also morally, protecting them with his colossal authority. Administrative and deputy duties left Vonsovsky little time for science, but the door to his office was always open. Talking to him was easy and simple: he was attentive to any person, regardless of his positions and regalia.

In the 1990s Vonsovsky gave more attention to humanitarian activities. He was one of the founders, rector and honorary president of the Humanitarian University of Yekaterinburg, where he himself lectured. Sergey Vasilyevich devoted the last years of his life to work on his memoirs and the textbook \cite{book}. Initially Vonsovsky planned to write a textbook on natural sciences for the Humanitarian University. But the book quickly went beyond this: he felt the need to convey not only his knowledge of physics, but also the cultural traditions of his generation. Vonsovsky managed to present many complex concepts of modern physics in a fascinating form, but without lowering the level: fundamental symmetries, the problem of the ``grand unification'' of all fundamental interactions, the theory of supersymmetry and superstrings, not forgetting about solid state physics, materials science and magnetism.

In the last chapter of his textbook, Sergey Vasilyevich cited the  words of Andrey Sakharov, whom he treated with great human respect. They probably also express the scientific-philosophical and ethical position of Vonsovsky himself: ``I support the cosmological hypothesis which states that the development of the universe is repeated in its basic features an infinite number of times. In accordance with this, other civilizations, including more ``successful'' ones, should exist an infinite number of times on the “preceding” and the ``following'' pages of the Book of the Universe. Yet this should not minimize our sacred endeavors in this world of ours, where, like faint glimmers of light in the dark, we have emerged for a moment from the nothingness of dark unconsciousness of material existence. We must make good the demands of reason and create a life worthy of ourselves and of the goals we only dimly perceive.''

\section{Acknowlegments}
We are grateful to I. Mazin and S. Ovchinnikov for useful discussions. Authors appreciate support of the Ministry of Science and Higher Education of the Russian Federation (theme ``Quantum'' No. 122021000038-7).

{}
\end{document}